\documentclass[11pt,french,english,aps,prl,reprint,longbibliography]{revtex4-1}
\usepackage[T1]{fontenc}
\usepackage[latin9]{inputenc}
\usepackage[a4paper]{geometry}
\usepackage{xcolor}
\usepackage{pdfcolmk}
\usepackage{bm}
\usepackage{amstext}
\usepackage{graphicx}
\usepackage{setspace}
\usepackage{esint}
\PassOptionsToPackage{normalem}{ulem}
\usepackage{ulem}

\makeatletter

\providecommand{\tabularnewline}{\\}
\providecolor{lyxadded}{rgb}{0,0,1}
\providecolor{lyxdeleted}{rgb}{1,0,0}
\DeclareRobustCommand{\lyxadded}[3]{{\color{lyxadded}{}#3}}

\usepackage{braket}
\usepackage{graphicx}
\usepackage[euler]{textgreek}

\makeatother

\usepackage{babel}
\makeatletter
\addto\extrasfrench{%
   \providecommand{\fg}{\ifdim\lastskip>\z@\unskip\fi~\frqq}%
}

\makeatother
\begin{document}

\title{Study of a water-graphene capacitor with molecular density functional
theory}

\author{Guillaume Jeanmairet}

\affiliation{Sorbonne Université, CNRS, Physico-Chimie des Électrolytes et Nanosystèmes
Interfaciaux, PHENIX, F-75005 Paris, France}

\affiliation{Réseau sur le Stockage Électrochimique de l\textquoteright Énergie
(RS2E), FR CNRS 3459, 80039 Amiens Cedex, France}

\author{Benjamin Rotenberg}

\affiliation{Sorbonne Université, CNRS, Physico-Chimie des Électrolytes et Nanosystèmes
Interfaciaux, PHENIX, F-75005 Paris, France}

\affiliation{Réseau sur le Stockage Électrochimique de l\textquoteright Énergie
(RS2E), FR CNRS 3459, 80039 Amiens Cedex, France$\ $}

\author{Daniel Borgis}

\affiliation{PASTEUR, Département de chimie, École normale supérieure, PSL University,
Sorbonne Université, CNRS, 75005 Paris, France}

\affiliation{Maison de la Simulation, CEA, CNRS, Universit\'{e} Paris-Sud, UVSQ,
Universit\'{e} Paris-Saclay, 91191 Gif-sur-Yvette, France}

\author{Mathieu Salanne}

\affiliation{Sorbonne Université, CNRS, Physico-Chimie des Électrolytes et Nanosystèmes
Interfaciaux, PHENIX, F-75005 Paris, France}

\affiliation{Réseau sur le Stockage Électrochimique de l\textquoteright Énergie
(RS2E), FR CNRS 3459, 80039 Amiens Cedex, France}

\affiliation{Maison de la Simulation, CEA, CNRS, Université Paris-Sud, UVSQ, Universit\'{e}
Paris-Saclay, 91191 Gif-sur-Yvette, France}
\begin{abstract}
Interfacial molecular processes govern the performances of electrochemical
devices. Aqueous electrochemical systems are often studied using classical
density functional theory but with too crude approximations. Here
we study a water-graphene capacitor, improving the state of the art
by the following key points: 1) electrodes have a realistic atomic
resolution, 2) a voltage is applied and 3) water is described by a
molecular model. We show how the permittivity and the structure of
interfacial water change with voltage. The predicted capacitance agrees
with molecular dynamics simulations.
\end{abstract}
\maketitle
\begin{singlespace}
The interface between a solution and an electrode is a complex physicochemical
system in which both the liquid and the solid properties strongly
differ from their bulk ones. For example, X-ray adsorption experiments
coupled with ab-initio molecular dynamics study of the water-gold
interface revealed an altered structure with respect to the bulk \cite{velasco-velez_structure_2014}.
Applying a voltage between electrodes also impacts the local structure
and interfacial properties of the liquid \cite{velasco-velez_structure_2014,siepmann_influence_1995}.
In the case of an electrolytic solution, ions adopt a complex structure
at the electrode giving rise to an electrical double layer (EDL) \cite{merlet_electric_2014}.
\end{singlespace}

Since direct experimental measurements are difficult, a lot of our
knowledge on the liquid-electrode interface at the molecular level
comes from theories and simulations. The simulation of electrochemical
systems is further complicated by the necessity to account for the
applied potential difference between the two electrodes. In most studies,
the adopted strategy is to impose uniform charge densities of opposite
sign at each electrode, which is not equivalent to fixing potential.
Siepmann and Sprik \cite{siepmann_influence_1995,willard_water_2008}
proposed a more advanced methodology in which each electrode atom
bears a Gaussian charge which values are determined to fix the potential
to the desired value. It has been reported that using the fixed charge
method in Molecular Dynamics (MD) simulations causes a quick and non-physical
raising of the temperature while this is not observed with the fluctuating
charge method \cite{merlet_electric_2014,merlet_simulating_2012}.
While MD studies provide an efficient toolbox to study electrochemical
systems, they are numerically expensive especially when the fluctuating
charges method is used. As an example, the recent computation of the
capacitance of a device made of two amorphous carbon electrodes immersed
in a sodium chloride aqueous solution took several millions CPU hours
\cite{simoncelli_blue_2018}. 

Classical density functional theory (cDFT) is a computationally less
demanding alternative which has been intensively used to study electrochemical
systems \cite{hartel_fundamental_2015,hartel_structure_2017,henderson_density_2011,jiang_density_2011,jiang_solvent_2012,kong_molecular_2015,lian_capacitive_2017,lian_hydrophilicity_2016,lian_impurity_2017,velasco-velez_structure_2014,hartel_dense_2016}.
However, in these studies three approximations are made: 1) electrode
is modeled by a smooth hard wall with no atomic structure, 2) the
applied potential is mimicked by fixing opposite and uniform charge
distribution on electrodes and 3) water is described either by a hard-sphere
fluid \cite{jiang_solvent_2012,lian_capacitive_2017}, or by a dielectric
constant \cite{hartel_dense_2016,hartel_fundamental_2015,hartel_structure_2017,jiang_density_2011,lian_hydrophilicity_2016,lian_impurity_2017}.
These three approximations are quite drastic and limit the relevance
of simulations to describe realistic systems since: 1) the electrode
structure plays a key role in the capacitance \cite{merlet_highly_2013},
2) it has been shown using MD that fixing the charge instead of the
potential affects the structure of the adsorbed layer \cite{merlet_simulating_2012}
and 3) solvation effects also play a major role in the adsorption
of ions at the interface \cite{merlet_simulating_2012}. To address
those three points, we propose to use the molecular density functional
theory (MDFT) framework to simulate a capacitor consisting of two
graphene electrodes separated by pure water at fixed applied potential
difference. Graphene is described by an atomistic model while water
is modeled by the molecular force field SPC/E.

MDFT has been extensively detailed in previous work \cite{zhao_molecular_2011,jeanmairet_molecular_2013-1,ding_efficient_2017,jeanmairet_molecular_2019}
and we only briefly recall its basics. In MDFT, a liquid (here water)
is described by its density field $\rho(\bm{r},\bm{\Omega})$, which
measures the average number of solvent molecules with orientation
$\bm{\Omega}$ at a given position $\bm{r}$. For any external perturbation
(here the presence of electrodes), it is possible to write a unique
functional $F$ of the density. This functional is usually split into
the sum of three terms:
\begin{equation}
F[\rho]=F_{\text{id}}[\rho]+F_{\text{ext}}[\rho]+F_{\text{exc}}[\rho].\label{eq:F=00003DFid+Fexc+Fext}
\end{equation}
The ideal term $F_{\text{id}}$ corresponds to the usual entropic
term for a non-interacting fluid. The second term $F_{\text{ext}}$
is due to the external potential $V_{\text{ext}}$ of the electrodes
acting on the liquid:
\begin{equation}
F_{\text{ext}}[\rho]=\iint\rho\left(\bm{r},\bm{\bm{\Omega}}\right)V_{\text{ext}}(\bm{r},\bm{\Omega})d\bm{r}d\bm{\Omega}\label{eq:Fext}
\end{equation}
Here, $V_{\text{ext}}$ is the sum of a Lennard Jones potential $V_{\text{ext,LJ}}$
and of an electrostatic term $V_{\text{ext,e}}$ due to interactions
between partial charges of water molecules and the fluctuating charges
on the electrodes. Finally, the last term of Equation \ref{eq:F=00003DFid+Fexc+Fext}
represents the solvent-solvent contribution for which we employ the
most accurate expression for SPC/E \cite{ding_efficient_2017}, which
corresponds to the so-called hypernetted chain or equivalently bulk
reference fluid approximation.

The density $\rho_{\text{eq}}$ minimizing the functional of Equation
\ref{eq:F=00003DFid+Fexc+Fext} is the Grand Canonical equilibrium
solvent density of the liquid. This density $\rho_{\text{eq}}$ is
inhomogeneous and thus generates an inhomogeneous charge distribution 

\begin{equation}
\rho_{c}(\bm{r})=\iint\rho_{\text{eq}}(\bm{r}^{\prime},\bm{\Omega})\widetilde{\rho_{c}}(\bm{r}-\bm{r}^{\prime},\bm{\Omega})d\bm{r}^{\prime}d\bm{\Omega}.\label{Solvent charge distrib}
\end{equation}
where $\widetilde{\rho_{c}}(\bm{r},\bm{\Omega})=\sum_{i}q_{i}\delta(\bm{r}-\bm{r}_{i_{\bm{\Omega}}})$
is the charge distribution of a single water molecule located at the
origin with an orientation $\bm{\Omega}$ , $\delta$ denotes the
Dirac distribution, the sum runs over the solvent sites, $q_{i}$
is the charge of site $i$ and $\bm{r}_{i_{\bm{\Omega}}}$ is the
position of this site when the molecule has an orientation $\bm{\Omega}$.
This inhomogeneous charge distribution polarizes the two electrodes. 

To account for the polarizability of electrodes under a fixed potential
difference, we employ the method proposed by Siepmann and Sprik. Each
electrode atom $j$ bears a Gaussian charge distribution of fixed
width 0.505 $\textrm{Å}$ and magnitude $q_{j}$. These charges are
treated as additional degrees of freedom and are determined to ensure
that the electrostatic potential $V_{j}$ experienced by each carbon
atom equals a prescribed value $V_{j,0}=\pm V_{0}$. In practice this
is done similarly to previous work \cite{siepmann_influence_1995,willard_water_2008},
by minimizing the functional 
\begin{equation}
E_{\text{e}}^{\text{tot}}=E_{\text{e}}-\sum_{j}q_{j}V_{j,0},\label{eq:Minimise to compute charge}
\end{equation}
 with respect to the charges $q_{j}$ where $E_{\text{e}}$ is the
electrostatic energy. This set of electrode charges $q_{j}$ creates
an electrostatic field $V_{\text{ext,e}}$ in the external potential
of Equation \ref{eq:Fext}, which in turn modifies the polarization
of the liquid. 

\begin{figure}
\centering{}\includegraphics[width=0.4\textwidth]{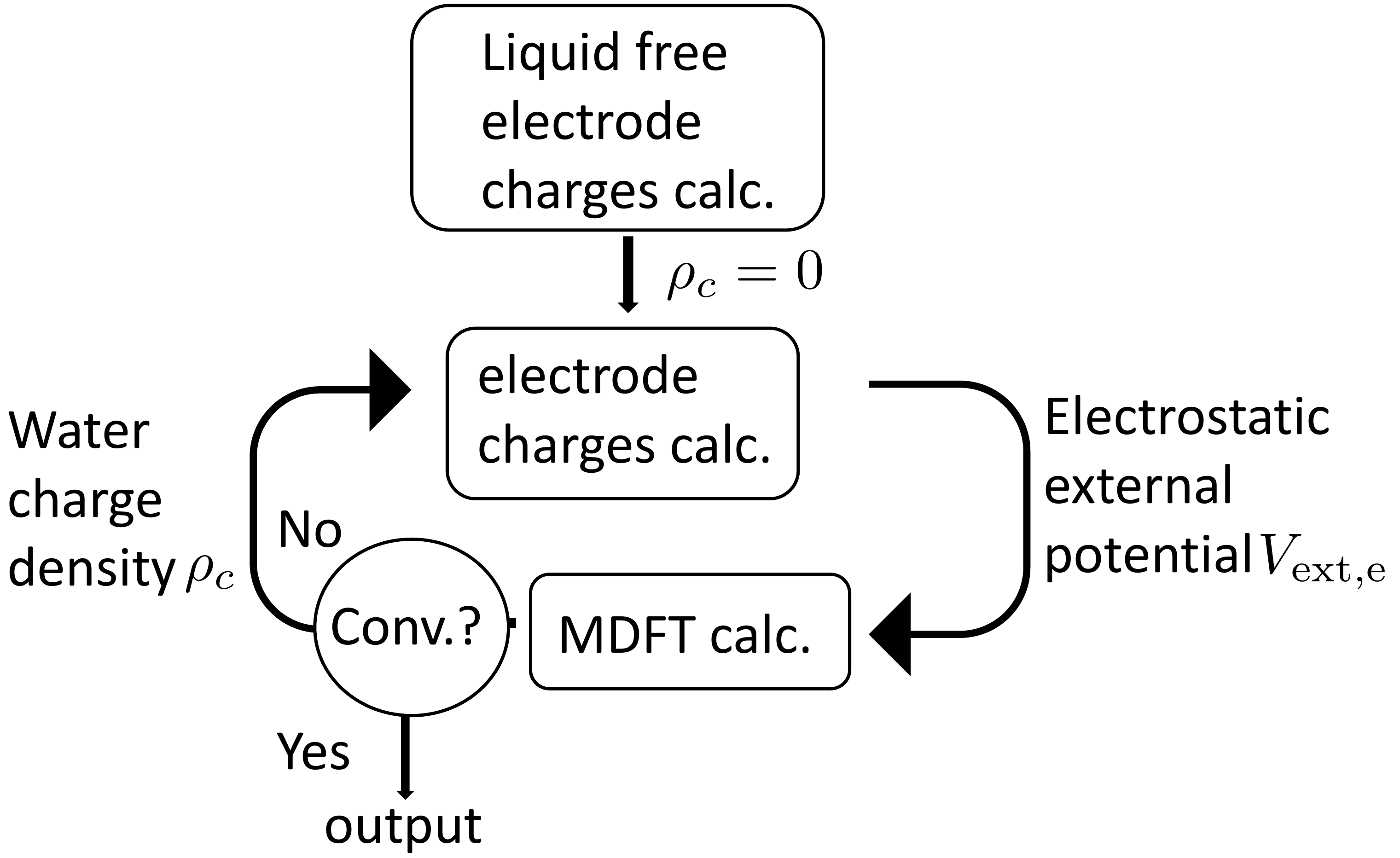}\caption{Workflow to compute the equilibrium solvent density and the charge
distribution within the electrodes under a fixed potential difference\label{fig:Shematic}\lyxadded{Maximilien Levesque}{Sat Apr 13 17:50:45 2019}{.}}
\end{figure}

To obtain the equilibrium solvent density and the electrode charges
we perform successive functional minimizations and electrode charges
optimizations as schematized in Figure \ref{fig:Shematic}. Initially,
we optimize the charges in the absence of solvent and compute the
electrostatic potential $V_{\text{ext,e}}$ it induces on each node
of the MDFT grid. This potential is then added to the Lennard Jones
potential $V_{\text{ext,LJ}}$ of the electrode atoms to compute the
external potential $V_{\text{ext}}$ entering Equation \ref{eq:Fext}.
After minimization of the functional, the equilibrium solvent charge
density $\rho_{c}$ of water is obtained from Equation \ref{Solvent charge distrib}.
It contributes to the total electrostatic potential energy $E_{\text{e}}$
in Equation \ref{eq:Minimise to compute charge}, which is minimized
to compute a new set of electrode charges. This process is iterated
until a convergence criterium is reached, typically when the relative
total charge variation between two iterations is below $5.0\ 10^{-4}$.

We study an electrochemical cell made of two graphene electrodes separated
by pure water. Each electrode made of 112 fixed carbon atoms has a
surface area of $17.36\times17.18\ \textrm{Å}^{2}$. The two electrodes
are separated by $60\ \textrm{Å}$. We run two sets of simulations.
In the first one water is described explicitly using fixed-potential
Molecular Dynamics, while in the second one we use the fixed-potential
MDFT described above. We run simulations with applied potential differences
$\Delta V$ of 0.0, 0.54, 1.1, 1.6 and 2.2 V. The force field parameters
are common to both setup: water is described by the SPC/E rigid model
and carbon atoms are modeled by a Lennard-Jones site with $\sigma=3.37\ \textrm{Å}$
and $\epsilon=0.23\ $kJ.mol$^{-1}$ \cite{cole_interaction_1983}.
The parameters for carbon-water interaction are obtained using the
Lorentz-Berthelot mixing rules. Carbon atoms also bear fluctuating
Gaussian charges interacting elecrostatically with water.

The MD simulations are preformed in the canonical ensemble at 298~K
and the inter-electrode space is filled with 560 water molecules in
order to recover the density of the homogeneous fluid ($0.033\ \textrm{Å}^{-3}$)
at the center of the cell. Since MDFT is a grand-canonical theory,
the number of water molecules is not an input parameter of the simulation
but rather a result from it. The predicted number of water molecules
is $560\pm1$ for all values of the applied potential. In MD, the
time step is 1 fs and statistics are accumulated for at least 400
ps after 50 ps of equilibration. 

Periodic boundary conditions (PBC) are only applied in the $x$ and
$y$ directions, \textit{i.e }parallel to the electrodes. In MDFT,
the excess solvent-solvent term is computed through the use of discrete
Fourier transform implying a 3D spatial periodicity. To be consistent
with the 2D-PBC of the external potential we suppress the undesired
periodicity along the $z$-axis by doubling the box size in this direction
and imposing a vanishing density by adding an infinite external potential
when $z>60\ \textrm{Å}$.

We first examine the in-plane-averaged equilibrium densities of the
H and O sites of water along the $z$ direction, defined as
\begin{equation}
n_{\text{A}}(z)=\frac{1}{L_{x}L_{y}}\iint\frac{\rho_{\text{A}}(\bm{r},\bm{\Omega})}{\rho_{\text{A}}^{\text{bulk}}}dxdyd\bm{\Omega}
\end{equation}
where $L_{x}$ and $L_{y}$ are the dimensions of the box in the $x$
and $y$ directions, $\rho_{\text{A}}$ is the 3D density of oxygen
or hydrogen and $\rho_{\text{A}}^{\text{bulk}}$ is the corresponding
density in the bulk solvent. In MD this quantity can be computed through
binning of the particles positions.

\begin{figure}
\centering{}\includegraphics[width=0.5\textwidth]{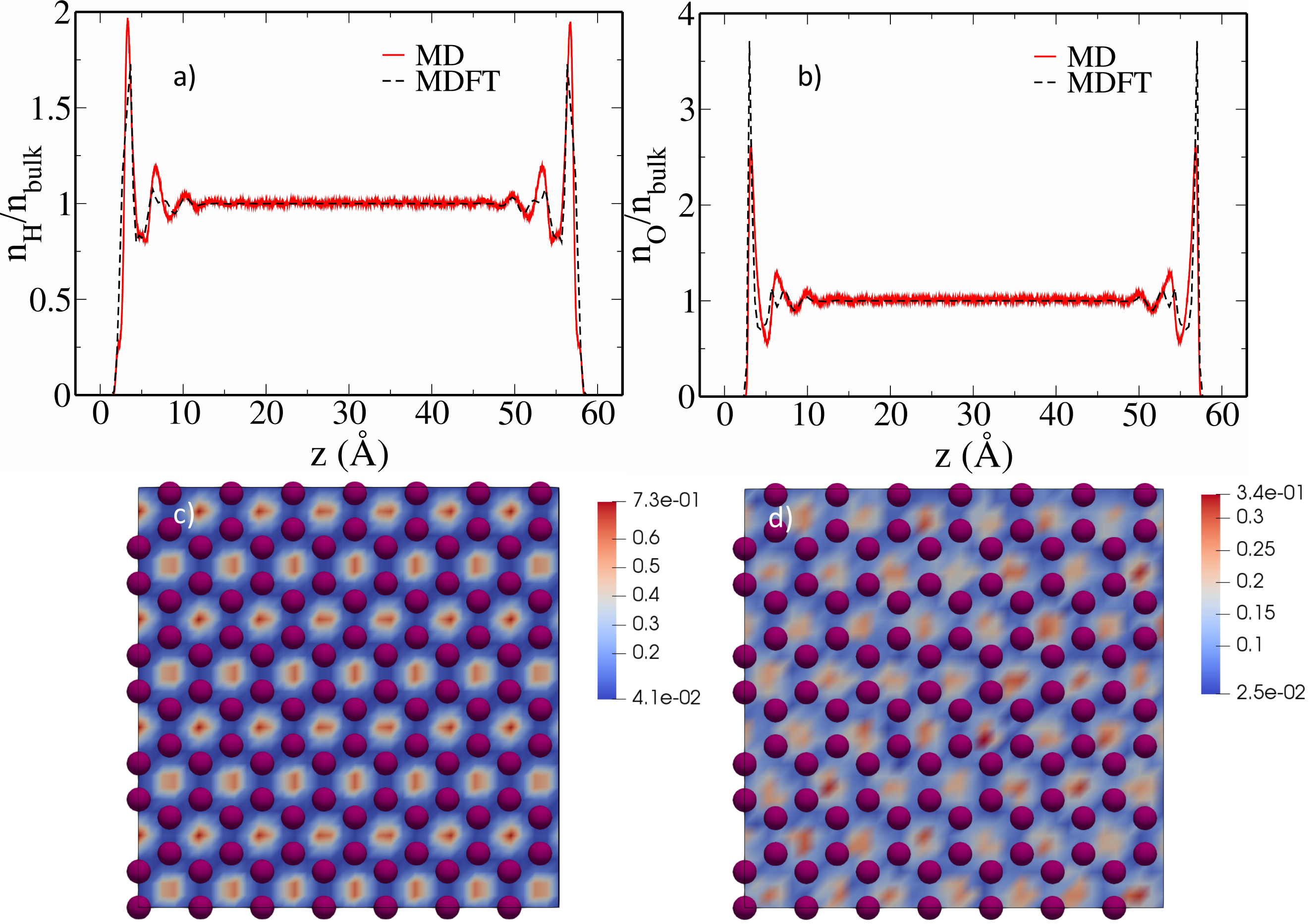}\caption{Top: In-plane-averaged equilibrium densities of H (a) and O (b) of
water along the $z$ direction. The MD results are in full red and
the MDFT results are in dashed black. Bottom: Slices of oxygen density
computed by MDFT (c) and MD (d) located at $z\approx2.5\ \textrm{Å}$,
the purple balls represent the graphene carbon atoms. \label{fig:densities}}
\end{figure}

Figure \ref{fig:densities} compares $n_{\text{H}}$ and $n_{\text{O}}$
computed with MD and MDFT for $\Delta V=0.0\ \text{V}$ (oxygen densities
for other values of applied voltage are given in supplementary materials).
Both density profiles are symmetric with respect to the center of
the cell since the two electrodes are identical; applying a non-zero
potential difference breaks this symmetry\textcolor{blue}{{} }\textcolor{black}{as
illustrated in SM.} The two methods give results in qualitative agreement
with two marked solvation peaks beyond which the homogeneous densities
are quickly recovered. A closer look reveals some discrepancies. For
the oxygen density in Figure \ref{fig:densities}.a, the maximum of
the first peak is located at $3.0\ \textrm{Å}$ with MDFT, slightly
closer to the electrode than with MD ($3.2\ \textrm{Å}$). This first
pick is also sharper in MDFT than in MD. These results are consistent
with previous work showing that MDFT tends to predict over-structured
solvation shells around hydrophobic solutes \cite{jeanmairet_molecular_2013}.
For the hydrogen density profile in Figure \ref{fig:densities}.b,
the agreement is also qualitative since MDFT tends to predict slightly
less structured solvation shells: The two main peaks of the density
profile are smaller than in MD. 

Despite those differences the agreement between MD and MDFT is good,
as confirmed by the 3D densities. In Figure \ref{fig:densities} c-d
we present the oxygen densities in planes perpendicular to the electrode
located at $z\approx2.5\ \textrm{Å}$. Note that if the two figure
are extremely similar, the legend differ. Both methods predict an
higher density at the center of the hexagonal ring of graphene and
lower density on top of the carbon-carbon bonds. As evidenced in Figure
\ref{fig:densities}.c-d, the MDFT is well defined, while the MD densities,
computed after 9 ns of simulation would require more sampling to be
fully converged. The effect of the evolution of the density with the
sampling time is given in SM. 

\begin{figure}
\centering{}\includegraphics[width=0.4\textwidth]{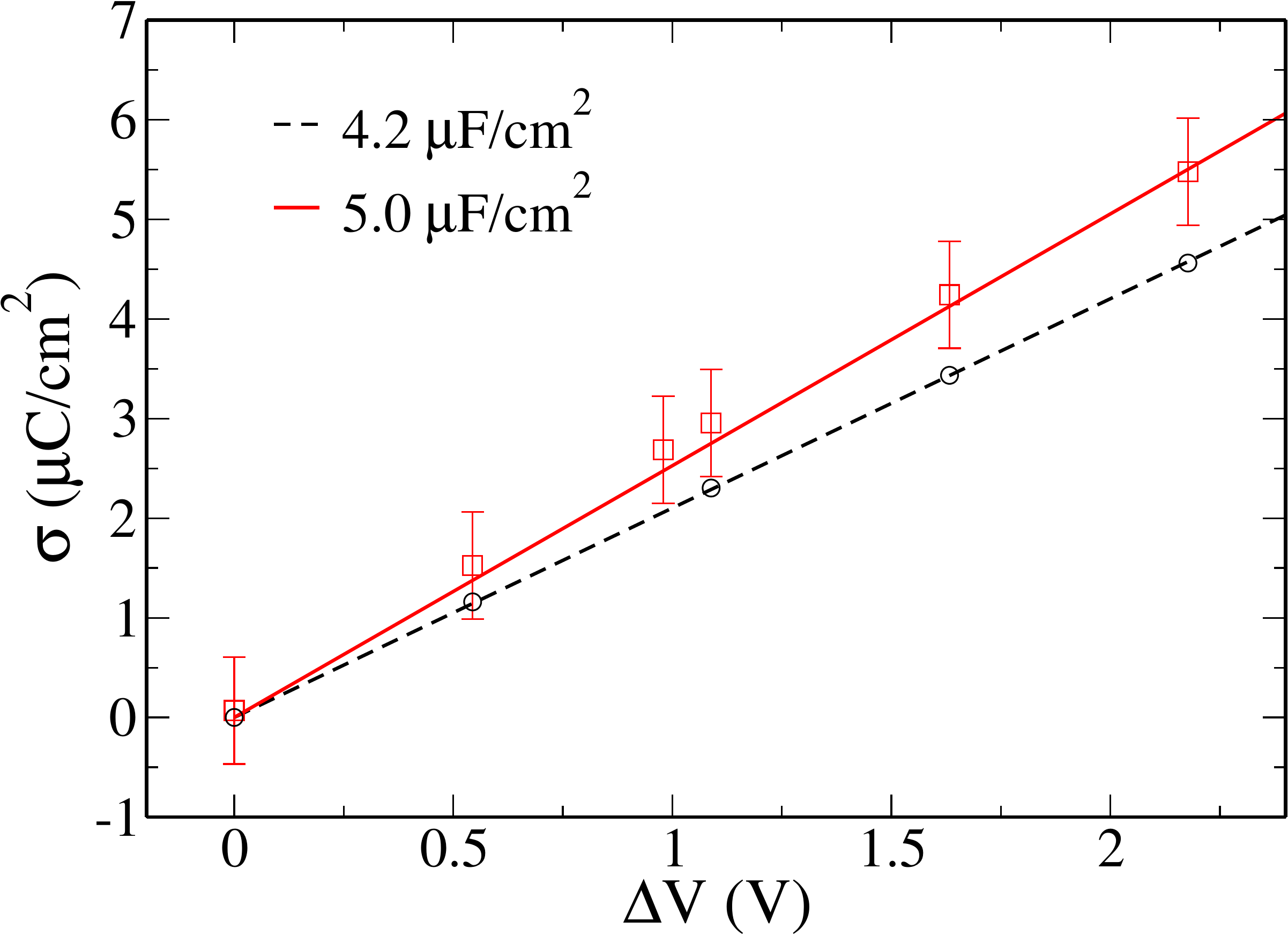}\caption{Surface charge density on the positive electrode as a function of
the applied potential difference. The MD results are the red square
and the error bars represent the standard deviation. MDFT results
are the black circles. The solid red line and the dashed black line
are linear fits of the data. The corresponding slopes are the capacitances
\label{fig:avQ}}
\end{figure}

We now turn to the performance of the water capacitor. Figure \ref{fig:avQ}
displays the charge density $\sigma$ on the positive electrode as
a function of the applied potential difference. Of course, the negative
electrode bears an opposite charge density. With both methods the
charge density varies linearly with the applied potential, which implies
a constant value of the differential capacitance over the range of
studied applied potentials. Capacitances are computed through linear
regression of the data, with the results $C=5.0\ $\textmu F.cm$^{-2}$
for MD and $C=4.2\ \text{}$\textmu F.cm$^{-2}$ for MDFT. 

\begin{figure}
\centering{}\includegraphics[width=0.4\textwidth]{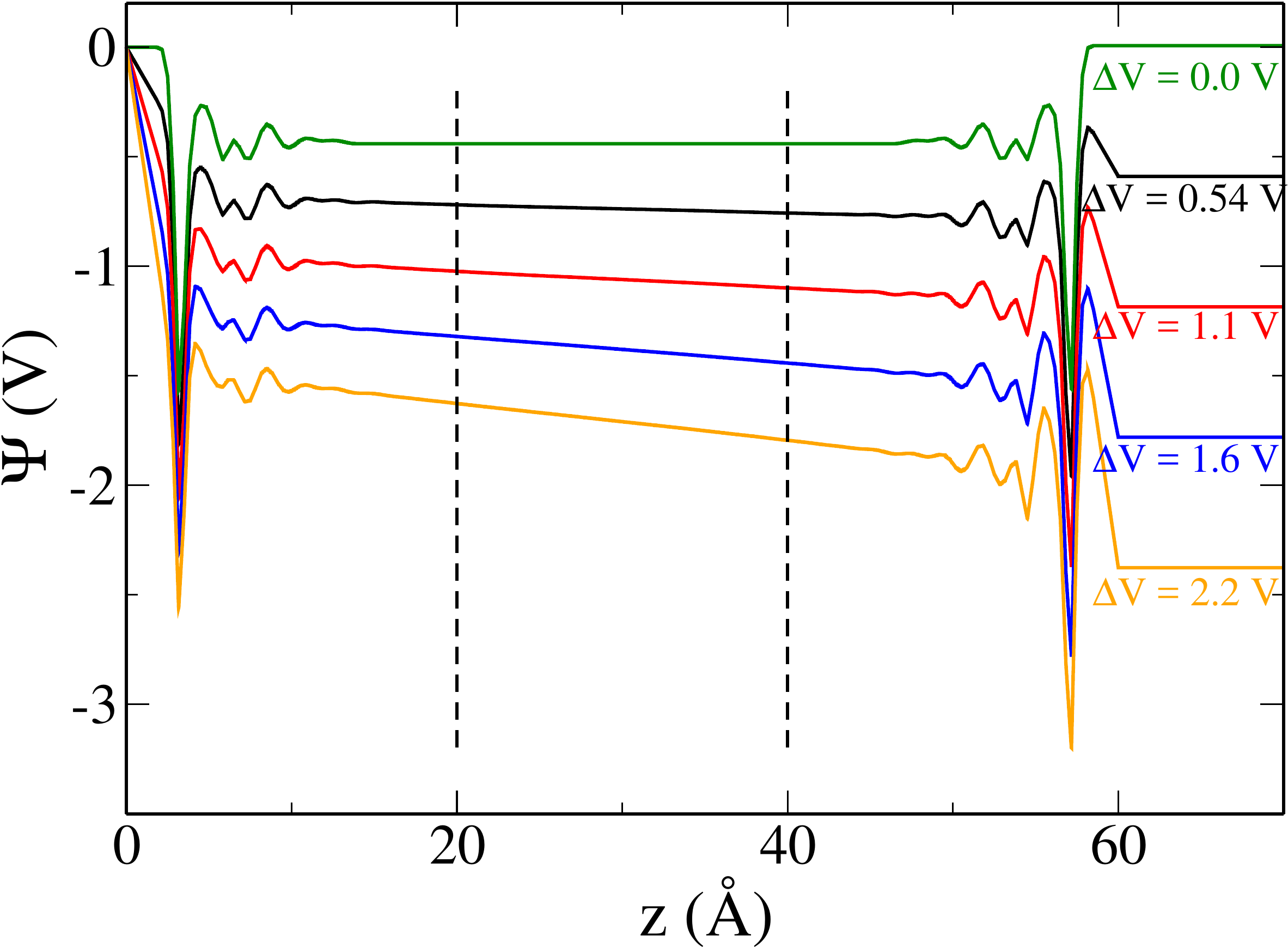}\caption{Poisson potential across the cell computed using Equation \ref{eq:Poisson}
for different values of the applied voltage $\Delta V$\label{fig:Poisson}}
\end{figure}

Knowing the charge distribution on each electrode, it is possible
to sum it with that of the solvent computed through Equation \ref{Solvent charge distrib}
to obtain the total charge distribution in the cell. This allows to
compute the evolution of the potential across the cell by solving
the Poisson Equation
\begin{equation}
\Delta\Psi(z)=-\frac{\bar{\rho_{c}}(z)}{\epsilon_{0}},\label{eq:Poisson}
\end{equation}
where $\bar{\rho_{c}}$ is the in-plane averaged total charge distribution
and $\epsilon_{0}$ is the vacuum permittivity. The Poisson potential
profiles computed with MDFT are displayed in Figure \ref{fig:Poisson}.
This potential is constant within the electrode. It then drops at
the interface and oscillates in a region of approximately 14 $\textrm{Å}$
with respect to the electrode before exhibiting a linear behavior
characteristic of a bulk dielectric material submitted to an external
electric field. The oscillation results from the layering of water
at the interface as observed in the density profiles of Figure \ref{fig:densities}.
Except for the short-circuit case ($\Delta V=0.0\ \text{V}$) the
potential is non symmetric because of different organizations of the
water molecules at the positive and negative electrodes, as is discussed
later. To compute the dielectric constant of water, we first calculate
the total charge density $\sigma_{L}$ (resp. $\sigma_{R}$) of adsorbed
water layers on the left (resp. right) electrode by integrating the
charge density of the solvent in the region of the cell $z<20\ \textrm{Å}$
(resp $z>40\ \textrm{Å}$). We then measure the potential drop $\Delta\Psi$
across the region between the planes at $z=20\ \textrm{Å}$ and $z=40\ \textrm{Å}$
in Figure \ref{fig:Poisson}. This allows to compute the surface capacitance
$C$ due to the dielectric liquid with the relation $\sigma_{L}=C_{S}\Delta\Psi$
for the different values of the applied potential. For a capacitor
composed of two planar metallic plates separated by a dielectric medium
the surface capacitance reads: 
\begin{equation}
C_{S}=\frac{\epsilon_{r}\epsilon_{0}}{d}\label{eq:C_capa_ideal}
\end{equation}
with $d$ the distance between the two plates and $\epsilon_{r}$
the permittivity of the dielectric medium. Results for the permittivity
$\epsilon_{r}$ are presented in Table \ref{tab:Dielectric-constant-of}
for different values of applied potential.

At low voltage, we obtain a value of $\epsilon_{r}=68.1$ in agreement
with previous values computed by MD by several groups \cite{yeh_dielectric_1999,apol_statistical_2002,zhang_computing_2016}.
When the applied potential increases, the computed dielectric constant
decreases, which is a known effect due to saturation of the dielectric
material \cite{yeh_dielectric_1999,willard_water_2008}. 

\begin{table}
\centering{}%
\begin{tabular}{|c|c|}
\hline 
$\Delta V$ (V) & $\epsilon_{r}$\tabularnewline
\hline 
\hline 
0.54 & 68.1\tabularnewline
\hline 
1.1 & 67.3\tabularnewline
\hline 
1.6 & 63.0\tabularnewline
\hline 
2.2 & 60.3\tabularnewline
\hline 
\end{tabular}\caption{Dielectric constant of SPC/E water computed for different values of
the applied voltage $\Delta V$.\label{tab:Dielectric-constant-of}}
\end{table}

\begin{figure}
\centering{}\includegraphics[width=0.4\textwidth]{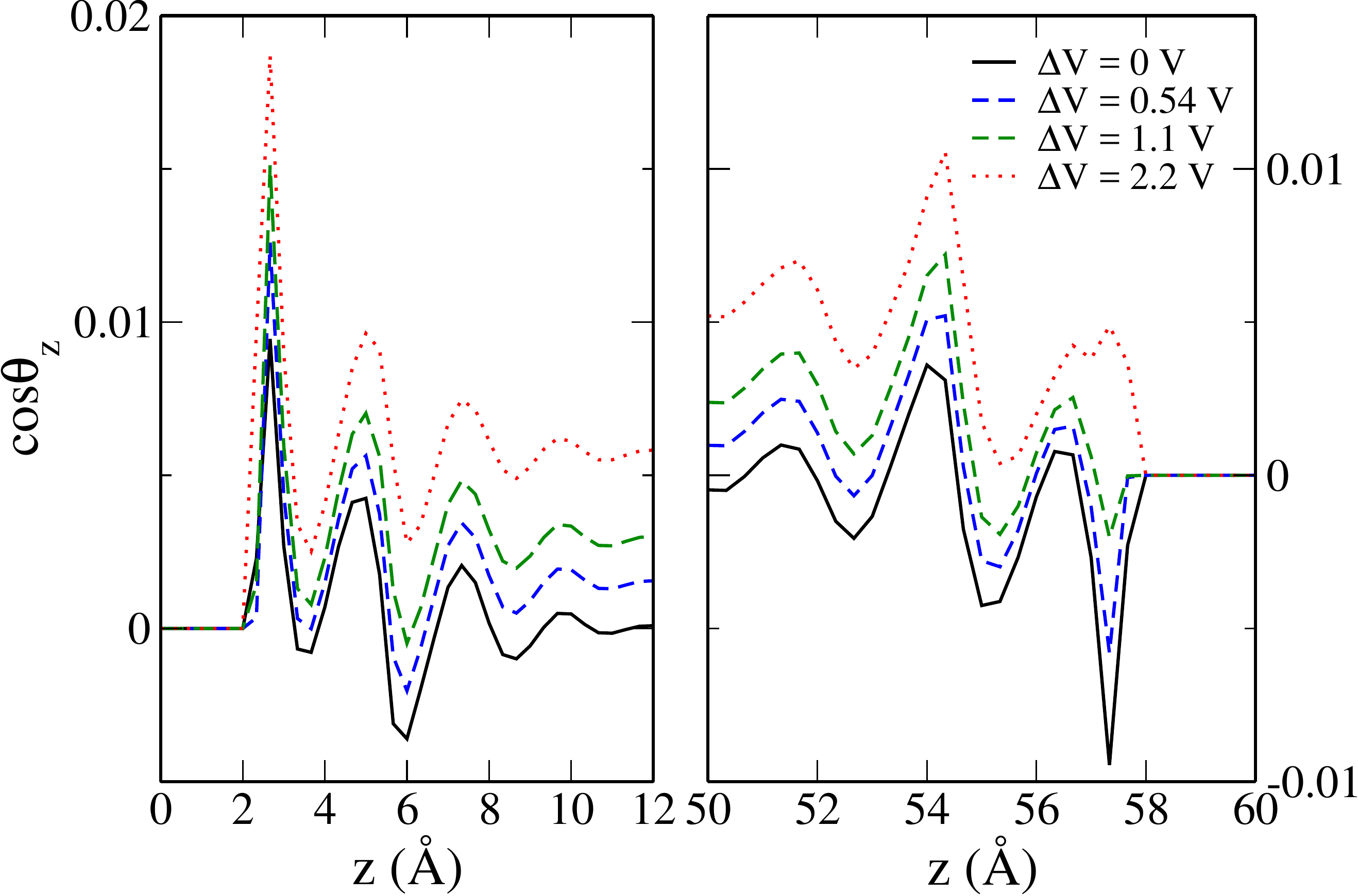}\caption{Average molecular polarization $\cos\theta_{z}$ along the normal
to the electrode surface.\label{fig:Pola_along_z}}
\end{figure}

Looking more closely at the evolution of the Poisson potential across
the two interfaces, we observe different behaviors at the two electrodes.
On the positive electrode (left), increasing the applied voltage $\Delta V$
simply enhances the potential drop observed at $0.0\ V$. On the negative
electrode (right), we observe a change of sign in the evolution of
the Poisson potential across the interface as the voltage increases.
In order to understand the molecular origin of this phenomenon we
report in Figure \ref{fig:Pola_along_z}, the planar-averaged $z$
component of the molecular dipolar polarization of the solvent, defined
as:
\begin{equation}
\cos\theta_{z}=\iiint\frac{\rho(\bm{r},\bm{\Omega})\bm{\Omega}\cdot\bm{e}_{z}}{\rho(\bm{r},\bm{\Omega})}dxdyd\bm{\Omega}.\label{eq:costheta}
\end{equation}
With the dipolar moment of a water molecule $\mu_{0}$ added as a
prefactor, the numerator of Equation \ref{eq:costheta} corresponds
to the $z$-component of the water dipolar polarization field. Thus,
$\cos\theta_{z}$ gives information about the average orientation
of water molecules located at a given $z$: When $\cos\theta_{z}=1$
the dipole of water molecule, pointing from oxygen to the center of
mass of hydrogen, is aligned with the $z$ axis. There are clearly
some preferential orientation of water with respect to the $z$ axis
for all values of the applied potential. We computed similarly $\cos\theta_{x}$
and $\cos\theta_{y}$ but no preferential orientation was observed. 

At $\Delta V=0.0\ \text{V}$, $\cos\theta_{z}$ is symmetric with
respect to the center of the cell. As expected, water molecules organize
identically at the two sides of the cell, and the oxygen are lying
closer to the surface than the hydrogen. Note that a purely dipolar
model (such as the Stockmayer fluid) would not polarize. The observed
finite and structured polarization emerges from the geometrical structure
of the water molecules, leading to a density-polarization density
coupling, that is fully accounted for in our molecular DFT approach
\cite{jeanmairet_molecular_2016}.

When the potential difference is increased, we observe a different
behavior of the polarization on each electrode. On the positive electrode
(left), increasing the potential enhances the oscillation of polarization
with respects to $\Delta V=0.0\ \text{V}$ and the global orientation
of water molecules at the interface is not modified. On the negative
electrode (right), we observe a flip of sign in the first peak of
polarization for voltage larger than $1.1\ V$, accompanied with a
slight shift of the position of this maximum toward the surface. This
indicates that for large enough voltage the preferential orientation
of adsorbed molecules is modified, as they flip with the hydrogen
atoms pointing toward the surface.

In summary, this paper reports a MDFT study of an electrochemical
device. Compared to other classical DFT calculations of similar systems,
this work introduce three major improvements: 1) the electrodes are
described by a realistic atomistic model, 2) the calculations are
done at fixed potential difference and 3) water is described by the
realistic molecular SPC/E model. Our implementation was tested on
a graphene-water capacitor and compared to fixed potential MD for
thorough validation. The solvation structure is almost quantitatively
predicted by MDFT. The prediction of the capacitance of the device
is good and the computed dielectric permittivity of the SPC/E water
model for different values of the applied potential difference agrees
well with previous MD simulations. It was also possible to take advantage
of the 3D resolution of the equilibrium solvent density to gain precise
insight into the orientational order of water at the interface and
its evolution with the applied potential difference. In this paper,
functional minimization and the electrode charge optimization were
carried out independently and sequentially. It should be possible
to propose a functional form depending on both the solvent density
and the electrode charges. Minimizing such a functional would allow
to find the equilibrium solvent density and the equilibrium electrode
charge distribution self consistently. This is a current direction
of research. The limitations of MDFT in the straight HNC approximation
considered here, especially close to mildly charged entities, have
already been pinpointed and so-called « bridge » functionals going
beyond HNC have been suggested \cite{levesque_scalar_2012,jeanmairet_molecular_2013,jeanmairet_molecular_2015}.
Inclusion of those in the present problem is a direction of investigations.

\section*{Acknowledgments}

The authors acknowledge Dr Maximilien Levesque for fruitful discussions.
B.R. acknowledges financial support from the French Agence Nationale
de la Recherche (ANR) under Grant No. ANR-15-CE09-0013 and from the
Ville de Paris (Emergences, project Blue Energy). This project has
received funding from the European Research Council (ERC) under the
European Union\textquoteright s Horizon 2020 research and innovation
programme (grant agreement No. 771294). This work was supported by
the Energy oriented Centre of Excellence (EoCoE), Grant Agreement
No. 676629, funded within the Horizon 2020 framework of the European
Union.

\selectlanguage{french}%
\selectlanguage{english}%

\end{document}